\documentclass[%
reprint,
superscriptaddress,
 amsmath,amssymb,
 aps,
prb,
]{revtex4-2}

\usepackage{graphicx}
\usepackage{dcolumn}
\usepackage{bm}
\usepackage{soul}
\usepackage[dvipsnames]{xcolor}
\usepackage{xfrac}
\usepackage{natbib}
\usepackage{hyperref}
\usepackage{xfrac}
\hypersetup{
    colorlinks=true,
    linkcolor=blue,
    filecolor=blue
    urlcolor=blue,
    citecolor=blue,
}
\usepackage{mathrsfs}
\usepackage[nice]{nicefrac}
\usepackage{xfrac}
\usepackage{lipsum}

\begin{document}


\title{Absence of a bulk signature of a charge density wave in hard x-ray measurements of UTe$_2$}

\author{Caitlin S.~Kengle}
\affiliation{Department of Physics, University of Illinois, Urbana, Illinois 61801, USA}%
\affiliation{Materials Research Laboratory, University of Illinois, Urbana, Illinois 61801, USA}

\author{Dipanjan Chaudhuri}
\affiliation{Department of Physics, University of Illinois, Urbana, Illinois 61801, USA}%
\affiliation{Materials Research Laboratory, University of Illinois, Urbana, Illinois 61801, USA}

\author{Xuefei Guo}
\affiliation{Department of Physics, University of Illinois, Urbana, Illinois 61801, USA}%
\affiliation{Materials Research Laboratory, University of Illinois, Urbana, Illinois 61801, USA}

\author{Thomas A.~Johnson}
\affiliation{Department of Physics, University of Illinois, Urbana, Illinois 61801, USA}%
\affiliation{Materials Research Laboratory, University of Illinois, Urbana, Illinois 61801, USA}

\author{Simon Bettler}
\affiliation{Department of Physics, University of Illinois, Urbana, Illinois 61801, USA}%
\affiliation{Materials Research Laboratory, University of Illinois, Urbana, Illinois 61801, USA}

\author{Wolfgang Simeth}
\affiliation{Laboratory for Neutron and Muon Instrumentation, Paul Scherrer Institute, CH-5232 Villigen PSI, Switzerland}%
\affiliation{Physik-Institut, Universit\"{a}t Z\"{u}rich, CH-8057 Z\"{u}rich, Switzerland}
\affiliation{Los Alamos National Laboratory, Los Alamos, New Mexico 87545, USA}%

\author{Matthew J.~Krogstad}
\affiliation{Materials Science Division, Argonne National Laboratory, 9700 South Cass Avenue, Argonne, Illinois 60439, USA}

\author{Zahir Islam}
\affiliation{Advanced Photon Source, Argonne National Laboratory, 9700 South Cass Avenue, Argonne, Illinois 60439, USA}

\author{Sheng Ran}
\affiliation{Maryland Quantum Materials Center, Department of Physics, University of Maryland, College Park, Maryland, USA}
\affiliation{NIST Center for Neutron Research, National Institute of Standards and Technology, Gaithersburg, Maryland, USA}
\affiliation{Department of Physics, Washington University in St. Louis, St Louis, Missouri, USA}

\author{Shanta R.~Saha}
\affiliation{Maryland Quantum Materials Center, Department of Physics, University of Maryland, College Park, Maryland, USA}

\author{Johnpierre Paglione}
\affiliation{Maryland Quantum Materials Center, Department of Physics, University of Maryland, College Park, Maryland, USA}
\affiliation{Canadian Institute for Advanced Research, Toronto, Ontario, Canada, M5G 1M1}

\author{Nicholas P.~Butch}
\affiliation{Maryland Quantum Materials Center, Department of Physics, University of Maryland, College Park, Maryland, USA}%
\affiliation{NIST Center for Neutron Research, National Institute of Standards and Technology, Gaithersburg, Maryland, USA}

\author{Eduardo Fradkin}
\affiliation{Department of Physics, University of Illinois, Urbana, Illinois 61801, USA}%
\affiliation{Materials Research Laboratory, University of Illinois, Urbana, Illinois 61801, USA}

\author{Vidya Madhavan}
\affiliation{Department of Physics, University of Illinois, Urbana, Illinois 61801, USA}%
\affiliation{Materials Research Laboratory, University of Illinois, Urbana, Illinois 61801, USA}

\author{Peter Abbamonte}
\affiliation{Department of Physics, University of Illinois, Urbana, Illinois 61801, USA}%
\affiliation{Materials Research Laboratory, University of Illinois, Urbana, Illinois 61801, USA}

\date{\today}

\begin{abstract}
    The long-sought pair density wave (PDW) is an exotic phase of matter in which charge density wave (CDW) order is intertwined with the amplitude or phase of coexisting, superconducting order. Originally predicted to exist in copper-oxides, circumstantial evidence for PDW order now exists in a variety of materials. 
    Recently, scanning tunneling microscopy (STM) studies have reported evidence for a three-component CDW at the surface of the heavy-fermion superconductor, UTe$_2$, persisting below its superconducting transition temperature [\href{https://doi.org/10.1038/s41586-023-06005-8}{Aishwarya, et al.~\textit{Nature} \textbf{618}, 928–933 (2023)}; \href{https://doi.org/10.1038/s41586-023-05919-7}{Gu, et al.~\textit{Nature} \textbf{618}, 921–927 (2023)}; \href{https://doi.org/10.1038/s41467-024-48844-7}{LaFleur, et al.~\textit{Nat.~Comm.} \textbf{15}, 4456 (2024)}]. Here, we use hard x-ray diffraction measurements on crystals of UTe$_2$ at $T = 1.9$ K and $12$ K to search for a bulk signature of this CDW. Using STM measurements as a constraint, we calculate the expected locations of CDW superlattice peaks and sweep a large volume of reciprocal space in search of a signature. We fail to find any evidence for a CDW near any of the expected superlattice positions in many Brillouin zones. We estimate an upper bound on the CDW lattice distortion of $u_{max} \lesssim 4 \times 10^{-3} \mathrm{\AA}$. 
    Our results suggest that the CDW observed in STM is either purely electronic, somehow lacking a signature in the lattice, or is restricted to the material surface. 
\end{abstract}

\maketitle

\section{\label{sec:Intro} Introduction}
Strongly correlated electronic systems often exhibit complex phase diagrams characterized by multiple ordered phases in close proximity. 
These broken-symmetry phases often compete or cooperate with one another, and exotic ground states, including unconventional superconductivity, can emerge as a result \cite{Venditti2023CondMatt, Fradkin2015RevModPhys, Fernandes2019AnnRevCondMatt}. 
When the phases cooperate to form a single ground state, the orders are said to be intertwined. 
A pair density wave (PDW) is a canonical example of such intertwined order in which superconductivity intimately interacts with charge density wave (CDW) order, resulting in a spatial modulation of the superconducting order parameter \cite{Agterberg2020AnnRev, Fradkin2015RevModPhys,Berg2009,Berg2009b}. 
Evidence for a PDW in real materials is mixed, the most compelling case currently being the cuprates \cite{DaSilvaNeto2014Science, 2019_Edkins_Science}. 
It is therefore of great interest to search for evidence of PDW phases in other materials.

Recently, superconducitivty was discovered in the actinide heavy fermion compound, UTe$_2$ with a transition temperature, $T_c$, as high as $2$ K, and possible spin-triplet pairing \cite{Ran2019Science, Aoki2019JPSJ, Rosa2022CommMatt}. 
Evidence for an exotic pairing mechanism comes from its extremely high and strongly anisotropic upper critical field \cite{Ran2019Science}. Additionally, multiple re-entrant superconducting phases have been observed at high magnetic fields, termed ``Lazarus superconductivity'', \cite{Ran2019Science, Frank2024NatComm} and under hydrostatic pressure \cite{2020_Thomas_SciAdv}. The presence of a PDW may provide insight to the nature of the superconducting order parameter, a source of active debate in UTe$_2$ \cite{2021_Hayes_Science,Rosa2022CommMatt,Aoki2019JPSJ,2020_Jiao_Nature,Ran2019NatPhys,Ran2019Science,2021_Ran_npjQuantum,2023_Broyles_PhysRevLett,2021_Duan_Nature, 2021_Thomas_PhysRevB, 2023_Ajeesh_PhysRevX}. 

Scanning tunneling microscopy (STM) is an important method for detecting CDWs and PDWs on the surfaces of materials \cite{ Hamidian2016Nature,2023_Zhao_Nature, Ruan2018NatPhys}. 
This method was notably used to detect a Cooper pair CDW in Bi$_2$Sr$_2$CaCu$_2$O$_{8+\delta}$ \cite{Hamidian2016Nature}.  
Similar observations were recently made in UTe$_2$: using both normal and superconducting tips, charge order peaks were identified in the tunneling spectra, indicating the possible existence of a pair density wave phase below $T \approx 4.5$ K \cite{Aishwarya2023Nature, 2023_Gu_Nature, Lafleur2024Nature}.
The charge density wave and pair density wave have coincident wave vectors, though the PDW is shifted in phase by $\pi$ \cite{2023_Gu_Nature}. 
While these observations are compelling, it is important to corroborate them with a different, ideally bulk-sensitive, technique. 

X-ray scattering is the quintessential method for detecting and characterizing CDWs \cite{Gruner1988RevModPhys, gruner2018density}. 
X-ray diffraction measures the energy-integrated (i.e., ``equal time") density-density correlation function of a material, giving a direct measurement of the charge density of the bulk, complimentary to STM, which measures the wavefunction overlap of a tip and a surface in the presence of a local electric field. 
Many charge density waves have been discovered using x-ray diffraction in one-dimesional (1D) systems, \cite{Hodeau1978JourPhysC, Sato1985JourPhysC}, layered chalcogenides \cite{Burk1992Science, Ivashko2017SciRep}, and copper-oxides \cite{DaSilvaNeto2014Science,Zimmermann1998EPL, BlancoCanosa2014PRB, Blackburn2013PRL}, among many others. Here, we apply this method to investigate CDW behavior in UTe$_2$.

\section{STM Summary}
\begin{figure}[h!]
    \centering
    \includegraphics[width=1\linewidth]{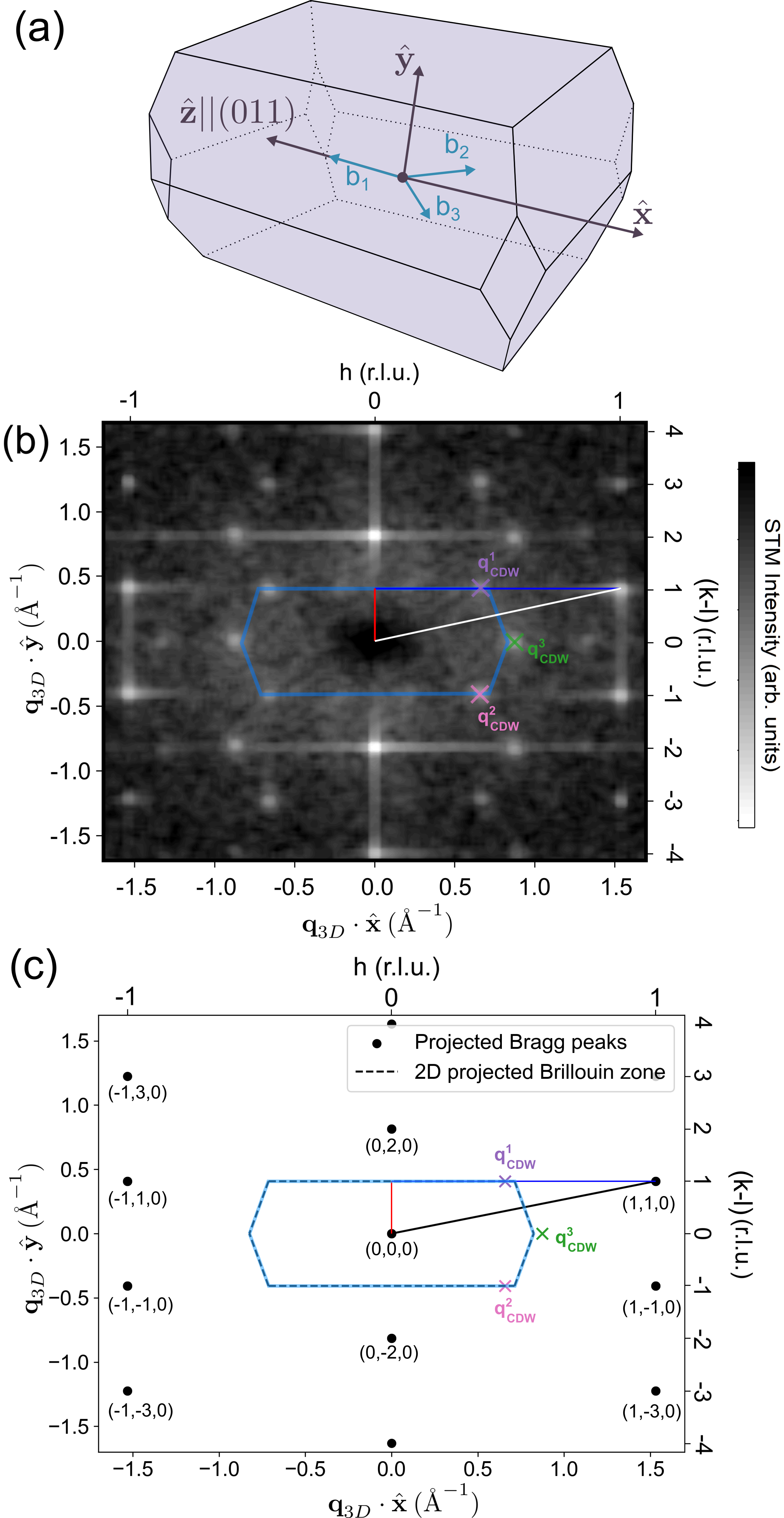}
    \caption{Relationship between STM measurements and 3D reciprocal space for UTe$_2$. (a) Body centered orthorhombic Brillouin zone. The blue arrows indicated by $b_{1,2,3}$ represent the primitive reciprocal lattice vectors. The black arrows represent the vectors discussed in Eqs.~\ref{eq:xdef}-\ref{eq:ydef}. (b-c) Surface projected Brillouin zone. (b) Zero bias STM measurement within the CDW phase (reproduced from \cite{Aishwarya2023Nature}), confirming designation of 2D projected Brillouin zone. (c) 3D Bragg peaks projected to [011] surface with the corresponding 2D surface projected Brillouin zone. Vector components between the $q=0$ point and the relevant Bragg peak and CDW positions according to Ref.~\cite{Aishwarya2023Nature}. The $kl$ values in the 3D Miller indices are not unique in this projection, so the labels are given as example.}
    \label{fig:STMsummary}
\end{figure}

UTe$_2$ has a proclivity to cleave along the (011) surface normal. 
It has an orthorhombic crystal structure (space group $Immm$).
The Brillouin zone of UTe$_2$, with the (011) direction indicated, is shown in Fig.~\ref{fig:STMsummary} (a).
A representative STM image of the (011) surface is shown in Fig.~\ref{fig:STMsummary} (b). The figure, reproduced from \cite{Aishwarya2023Nature}, shows the Fourier transform of the local density of states taken at the Fermi energy, $E_F$, at $T = 300$ mK. Three CDW modulations are visible in the data at incommensurate wavevectors, $q_1^{CDW}$, $q$$_2$$^{CDW}$, and $q_3^{CDW}$. These modulations, described in detail in Refs. \cite{Aishwarya2023Nature, 2023_Gu_Nature, Lafleur2024Nature}, appear below $T \approx 4.5$ K and possibly persist up to temperatures as high as $T \approx 10$ K. In the surface coordinates defined in Ref. \cite{Aishwarya2023Nature}, which we summarize in Fig.~\ref{fig:STMsummary}(c), the CDW wave vectors have coordinates $q_1^{\mathrm{CDW}} = (0.43q_x, q_y)$,  $q_2^{\mathrm{CDW}} = (0.43q_x, -q_y)$, and  $q_3^{\mathrm{CDW}} = (0.57q_x, 0)$, where $q_x$ and $q_y$ are the vector components of a nearby Bragg peak. Here we seek to see if these CDW modulations are present in the bulk of UTe$_2$ through low-temperature x-ray diffraction measurements.

\section{\label{sec:Exp} Experiment}
Experiments were done at beamline 6ID-D at the Advanced Photon Source at Argonne National Laboratory on a sample grown under the same conditions as those in \cite{Aishwarya2023Nature}. The incident x-ray energy was $E_i = 100$ keV (wavelength $\lambda = 0.124 \, \mathrm{\AA}$), at which the penetration depth in UTe$_2$ is $\sim 0.6$ mm, making this a bulk measurement. 
Using high energy x-rays has proven effective in detecting purely electronic CDWs, which occur without a significant periodic lattice distortion \cite{Zimmermann1998EPL}.

The samples were encapsulated in GE varnish to reduce exposure to air during transportation and throughout the experiment. 
The sample was loaded into a vacuum chamber with a transparent beryllium dome and a radiation shield held at a base pressure $\sim 10^{-5}$ Torr.
Sample temperature was controlled by an ARS DE302 three-stage cryocooler in which pre-cooled helium gas was fed into an open circuit Joule-Thomson stage to reach a base temperature of $T = 1.9$ K.
Data were collected using a Pilatus 3 x CdTe 2M detector at $T = 1.9$ and $12$ K.
Full, three-dimensional (3D) reciprocal space maps were performed by rotating the sample in the beam by 145$^\circ$ using a Huber four circle diffractometer.  

At $T = 1.9$ K, the lattice parameters were found to be $a = 4.099(20) \: \mathrm{\AA}$, $b = 6.137(28) \: \mathrm{\AA}$, and $c = 14.145(64) \: \mathrm{\AA}$. At $T=12$ K, the lattice parameters were found to be $a = 4.101(20) \: \mathrm{\AA}$, $b=6.135(28) \: \mathrm{\AA}$, and $c=14.147(65) \: \mathrm{\AA}$ Errors are the result of the optimization used to determine the lattice parameters. Deviations in lattice parameter values relative to other low-temperature reports \cite{Hutanu2020acta} arise as a result of sample mosaicity.

\begin{figure}
    \centering
    \includegraphics[width=0.99\linewidth]{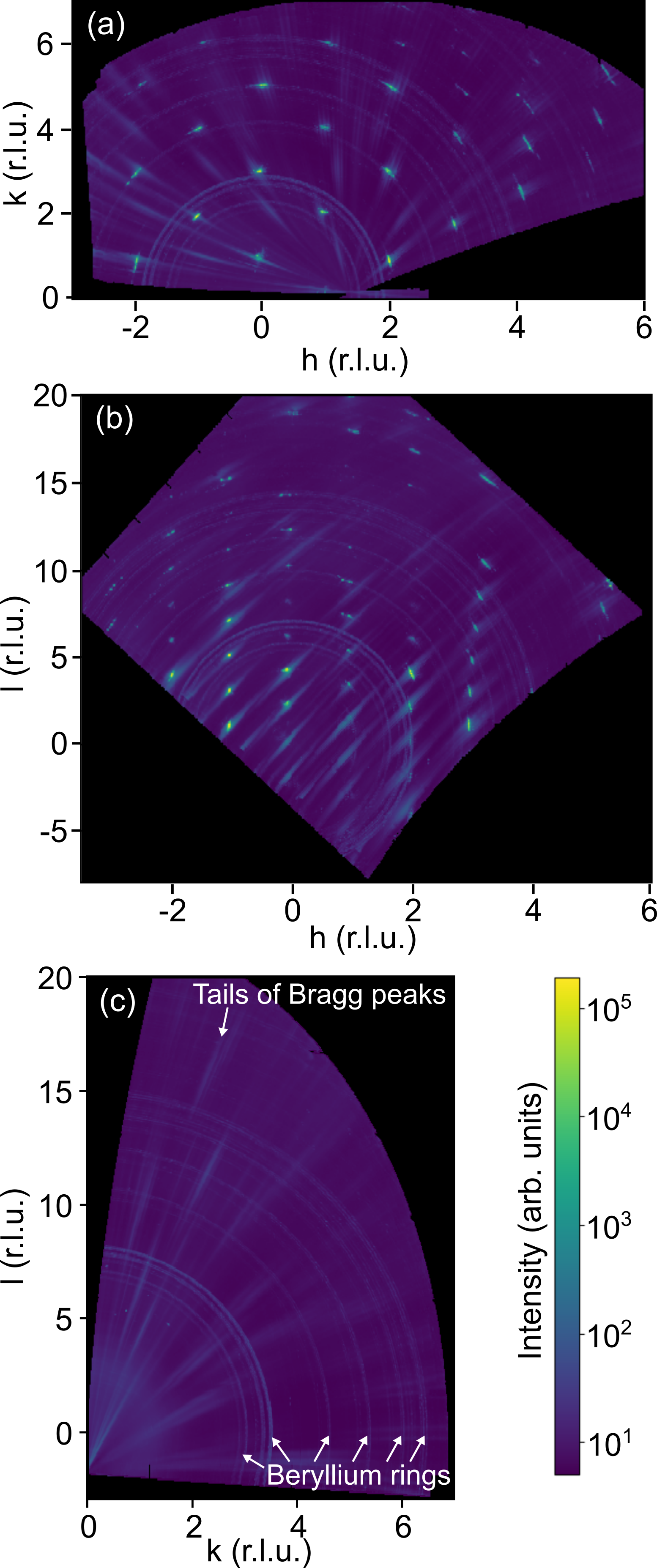}
    \caption{$T = 1.9$ K measurement of reciprocal space in UTe$_2$ crystal. (a) $hk$ cut along $l = 5$ r.l.u. and (b) $hl$ cut along $k = 2$ r.l.u. showing Bragg peak reflections obeying $h+k+l = 2n$. (c) $kl$ cut along non-integer $h = -0.43$ r.l.u. The weak signal comes from scattering from the beryllium domes and from the tails of Bragg peaks caused by the finite bandwidth of the x-ray source. All panels have the same intensity scale.}
    \label{fig:reciprocal_space}
\end{figure}

A major advantage of high energy x-rays is that they give access to a large volume of momentum space. Representative 2D cuts through 3D reciprocal space are shown in Fig.~\ref{fig:reciprocal_space} (a) with $l=5$ and (b) with $k = 2$. 
Sharp Bragg peaks at integer Miller indices, $(h,k,l)$, illustrate the crystalline quality of the samples. Figure \ref{fig:reciprocal_space} (c) shows a scan of the $kl$ plane taken at non-integer $h=-0.43$ r.l.u. at $T = 1.9$ K, which shows the primary background features: scattering from the beryllium dome and radiation shield, and streaking from Bragg peaks caused by the finite bandwidth of the incident x-ray beam. These features were found to be temperature-independent from our base temperature of T = 1.9 K to higher temperature measurements at T = 12 K, which is above the CDW phase transition observed in Refs.~\cite{Aishwarya2023Nature,2023_Gu_Nature, Lafleur2024Nature}.
No sign of a CDW was observed at any of the cuts of three-dimensional reciprocal space in Fig.~\ref{fig:reciprocal_space}.

\begin{figure*}[t!]
    \centering
    \includegraphics[width=1\linewidth]{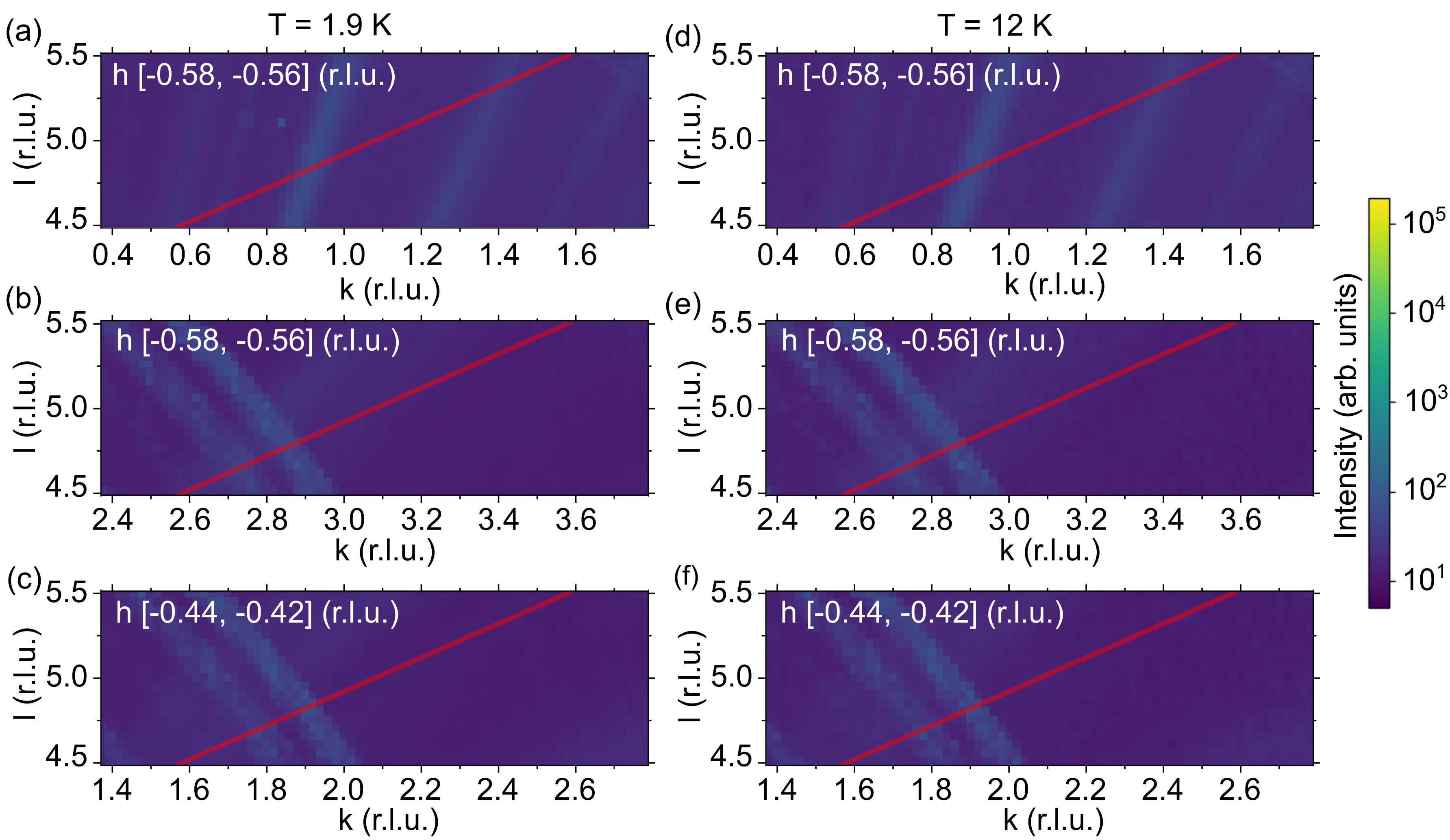}
    \caption{Diffraction images taken below and above the CDW transition temperatures in the $k,l$ plane at the appropriate $h$ values. The red line represents the calculated CDW positions in this zone for (a) $q^{CDW}_1$, (b) $q^{CDW}_2$, and (c) $q^{CDW}_3$ at $T= 1.9$ K. To compare, the calculated positions of (d) $q^{CDW}_1$, (e) $q^{CDW}_2$, and (f) $q^{CDW}_3$ at $T= 12$ K, above the CDW transition temperature, are also shown. Streaks in (a) and (d) come from the finite bandwidth of the incident x-ray beam. Rings in (b),(c),(e), and (f) come from scattering from amorphous beryllium of the sample enclosure. All images have the same intensity scale.}
    \label{fig:CDWPositions}
\end{figure*}

\section{Calculation of CDW Positions}
We will now focus on the regions of reciprocal space where the CDW is expected to be visible, based on what is known from the STM measurements of Refs.~\cite{Aishwarya2023Nature,2023_Gu_Nature, Lafleur2024Nature}. Because STM is a surface measurement, it does not constrain the periodicity of the CDW in the direction perpendicular to the surface. However, we can use the STM results to place partial constraints on the Miller indices, and identify what regions of momentum space the CDW is expected to be visible, assuming that the wave vectors at the surface are the same as in the bulk.

As noted before, all measurements of the CDW in STM are on the cleaved $(011)$ surface. In sample coordinates, the unit vector normal to this surface  $\hat{\textbf{z}} = \nicefrac{\overrightarrow{\textbf{z}}}{|\textbf{z}|} = \nicefrac{\left(0, c, b\right)}{\sqrt{b^2 + c^2}}$. The two vectors orthonormal to $\hat{\textbf{z}}$, which define the coordinate system of the surface, are
\begin{subequations}
    \begin{align}
        \hat{\textbf{x}} &= \left(1, 0, 0\right), \label{eq:xdef} \\    
        \hat{\textbf{y}} &= \hat{\textbf{z}} \times \hat{\textbf{x}} \label{eq:ydef},
    \end{align}
    \label{eq:2dsurfcecoordinates}
\end{subequations}
where $\hat{\textbf{x}}$ is defined to coincide with the $(h,k,l) = (1,0,0)$ crystallographic direction.

The peaks observed in an STM measurement [e.g., Fig.~\ref{fig:STMsummary}(b)] are projections of the bulk 3D Bragg peaks, which are indexed by Miller indices $(h,k,l)$, onto to the $(x,y)$ plane defined by Eqs.~\ref{eq:xdef} and \ref{eq:ydef}. The projection is given by the two-component vector
\begin{align}
    {q}_{2D} &= \biggl(\hat{\textbf{x}}\cdot \textbf{{q}}_{3D}, \hat{\textbf{y}}\cdot \textbf{{q}}_{3D}\biggr)
    \label{eq:3dprojection}
\end{align}
where $\textbf{q}_{3D} = \biggl(h\dfrac{2 \pi}{a}, k\dfrac{2 \pi}{b}, l\dfrac{2 \pi}{c}\biggr)$. The locations of these projected points, and the associated 2D  Brillouin zone of the surface, are shown in Fig.~\ref{fig:STMsummary}(c) and are consistent with STM data in Fig.~\ref{fig:STMsummary}(b). (Note that we have included only peaks with $h+k+l=2n$, where $n$ is an integer, the others being forbidden by crystal symmetry). 
In terms of the surface basis vectors Eq.~\ref{eq:2dsurfcecoordinates}, the CDW modulations observed in Ref.~\cite{Aishwarya2023Nature}, are  ${q}^{\mathrm{CDW}}_1 = \biggl(0.43\dfrac{2 \pi}{a}, \dfrac{2 \pi}{\sqrt{b^2 + c^2}}\biggr)$, ${q}^{\mathrm{CDW}}_2 = \biggl(0.43 \dfrac{2 \pi}{a}, \dfrac{-2 \pi}{\sqrt{b^2 + c^2}}\biggr) $, and ${q}^{\mathrm{CDW}}_3 = \biggl(0.57\dfrac{2 \pi}{a}, 0\biggr)$.

The 3D positions of each of the three CDW modulations, $q_i^{CDW}$, such as can be constrained by a 2D STM measurement, are determined by equating,
\begin{equation}
    q_{2D}=q_i^{CDW}.
    \label{eq:2dto3d}
\end{equation}
Doing so, we find that the STM experiment fully constrains the $h$ components of each of the three CDWs, i.e., 

\begin{subequations}
    \begin{align}
           h^{CDW}_1 &= 0.43 \: \mathrm{r.l.u.}  \label{eq:q1_3d_component}
            \\  
            h^{CDW}_2 &= 0.43 \: \mathrm{r.l.u.} \label{eq:q2_3d_component}
             \\
            h^{CDW}_3 &= 0.57 \: \mathrm{r.l.u.} \label{eq:q3_3d_component}.
    \end{align}
    \label{eq:q_3d_component}
\end{subequations}
The $k$ and $l$ components cannot be determined independently, but Eq.~\ref{eq:2dto3d} establishes a relationship between them:

\begin{subequations}
    \begin{align}
        k^{CDW}_1  &= l^{CDW}_1 - 1  \: \mathrm{r.l.u.}  \label{eq:qcdw1}
            \\  
        k^{CDW}_2 &=  l^{CDW}_2 + 1\: \mathrm{r.l.u.}  \label{eq:qcdw2}
             \\
        k^{CDW}_3 &= l^{CDW}_3  \label{eq:qcdw3}    
    \end{align}
    \label{eq:q_3d_component_kl}
\end{subequations}
Note that, unlike Eq.~\ref{eq:q_3d_component}, Eq.~\ref{eq:q_3d_component_kl} contain only natural numbers. This suggests, though is not proof, that the incommensurate character of the CDW in UTe$_2$ occurs only in the $(1,0,0)$ reciprocal lattice direction. 

Combining Eq.~\ref{eq:q_3d_component} and \ref{eq:q_3d_component_kl}, we find the expected locations of the three CDW wavevectors in UTe$_2$ derived from STM measurements to be $\textbf{q}^{CDW}_1 = (0.43, l^{CDW}_1-1, l^{CDW}_1)$, $\textbf{q}^{CDW}_2 = (0.43, l^{CDW}_2+1, l^{CDW}_2)$, and $\textbf{q}^{CDW}_3 = (0.57, l^{CDW}_3, l^{CDW}_3)$, where the values of $l^{CDW}_i$ are unknown. These expressions constrain the locations of the CDWs to one-dimensional cuts through 3D reciprocal space. Note that these vectors are defined relative to any crystalline Bragg peak, implying a large number of symmetry-related CDW satellites that could potentially be visible in any Brillouin zone. 

Note that, as far as zero-field measurements are concerned, all three of these modulations can be expressed in terms of single wave vector, $\textbf{q}^i_{3D}$, with the CDW being visible in different Brillouin zones with wave vector given by $\textbf{Q}_{CDW} = \textbf{G}_{3D} + \textbf{q}^i_{3D}$, with $\textbf{G}_{3D}$ being a reciprocal lattice vector. These different momentum-space points only become inequivalent in an applied magnetic field \cite{Aishwarya2023Nature}.

\section{Results}\label{sec:results}
The wide reciprocal space cut shown in Figure \ref{fig:reciprocal_space} (c), done at fixed $h=-0.43$, should intersect the $q^{CDW}_1$ and $q^{CDW}_2$ positions from the $h = 0$ zone as well as the $q^{CDW}_3$ position from the $h=-1$ zone. However, no CDW signal is observed above the experimental background at any of the three expected locations.

With hopes of uncovering a more subtle effect, we performed temperature difference measurements, comparing the same regions of reciprocal space at $T=1.9$ K and $T=12$ K, the latter being above the putative CDW ordering temperature \cite{Aishwarya2023Nature, 2023_Gu_Nature, Lafleur2024Nature}. We focused on three Brillouin zones near $(-1,1,5)$, $(-1,3,5)$, and $(-1,2,5)$, shown in Fig.~\ref{fig:CDWPositions}. These data sets were integrated over a narrow range of $h$ indicated in square brackets in each panel. The red lines indicate the possible CDW locations constrained from Eqs.~\ref{eq:q_3d_component} and Eqs.~\ref{eq:q_3d_component_kl} in each region. No changes with temperature were observed in any of these three Brillouin zones, or any of the other Brillouin zones mapped in Fig.~\ref{fig:reciprocal_space}.  
\begin{figure}[t!]
    \centering
    \includegraphics[width=1\linewidth]{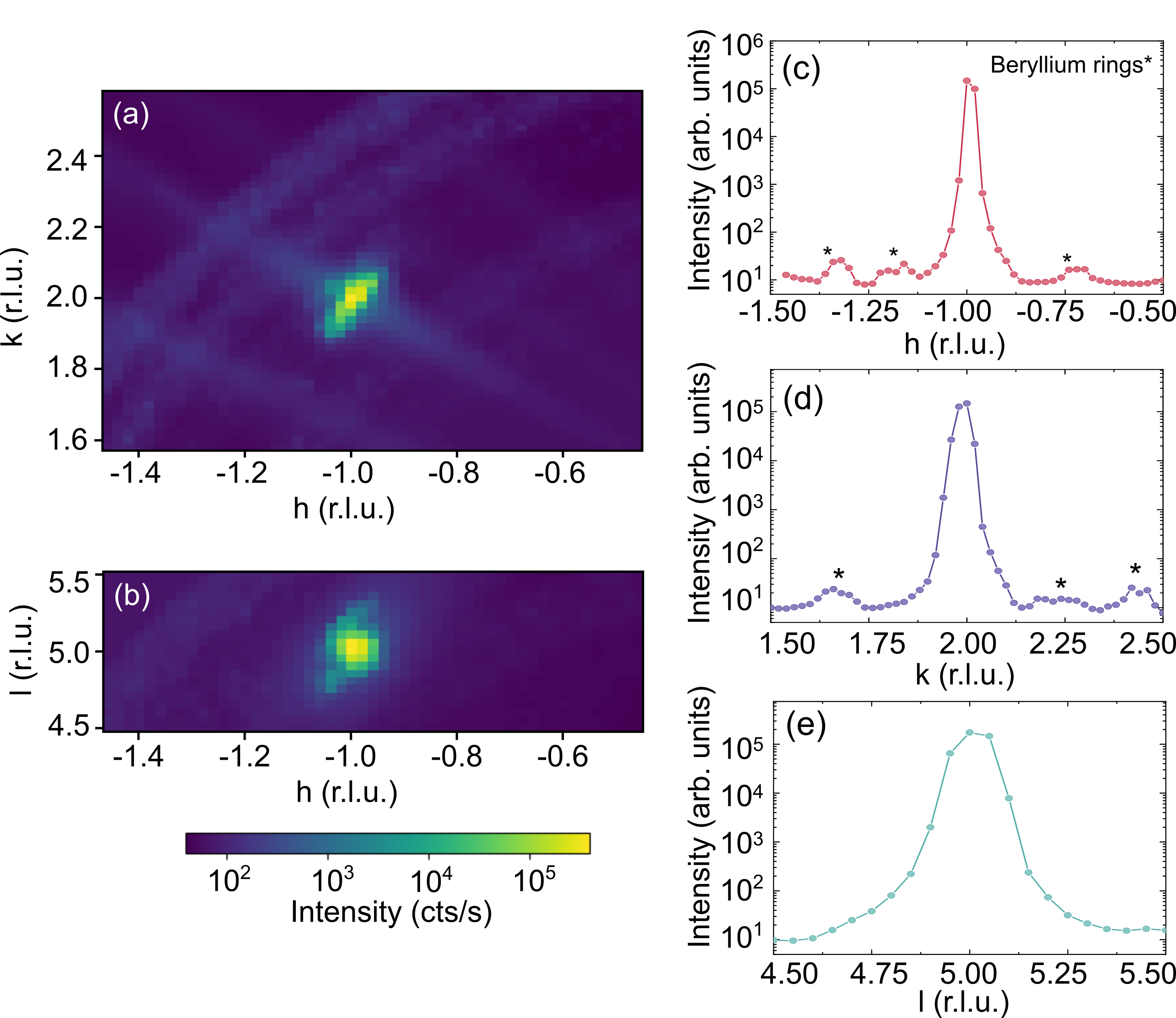}
    \caption{Scan through (-1,2,5) Bragg peak at $T = 1.9$ K in the (a) $hk$ plane and (b) $hl$ plane. Both plots are integrated $\pm 0.05$ r.l.u. in the out-of-plane direction. Line cut through Bragg peak along the (c) $h$, (d), $k$, and (e) $l$ directions.}
    \label{fig:-125_1p9K_BraggPeak_fits}
\end{figure}

Lacking a clear signature of a CDW, we can nevertheless use our data to restrict the maximum CDW amplitude in UTe$_2$. The background level of our measurements places an upper limit on the CDW peak intensity. The majority of the background comes from the beryllium rings (clearly visible in the $hk$ plane and line cuts), whose average photon counting intensity is $\sim 10$ Hz in this reciprocal space region. Assuming the photon noise is Poisson-distributed, this places an upper bound on the maximum CDW intensity of $I_{CDW}^{max}\leq \sqrt{10} \approx 3$ Hz.

How this intensity translates into a bound on the CDW amplitude depends on the CDW correlation length. 
In the STM data, the CDW peaks are of similar width to the crystalline Bragg peaks (Fig.~\ref{fig:STMsummary}(b)) whose width, in turn, is limited by the field of view of the STM image. The CDW width may also be field-of-view limited, or it may be inherently that broad. 
We therefore consider two cases: (1) The intrinsic width of the CDW is as observed in the STM measurements, and (2) the CDW has the same width as the structural Bragg peaks observed in our x-ray measurements. The latter case is summarized in Fig.~\ref{fig:-125_1p9K_BraggPeak_fits}.

The square of the amplitude of the CDW is proportional to the integrated intensity of the CDW reflection, which can be estimated,
\begin{equation}
    I_{CDW}^{Integrated} \lesssim \frac{1}{2}(I_{CDW}^{max})(f_{CDW})^3 
    \label{eq:integrated_intensity}
\end{equation}
where $f_{CDW}^3$ is the three-dimensional width of the CDW in reciprocal space. In the case where the CDW width is taken to be the same as in STM measurements, $f_{CDW}^3= (\xi^{-1}_{STM})^3$, where $\xi^{-1}_{STM}$ is the full-width at half-maximum of the CDW peak in STM, measured along the $q_x$ direction in Fig.~\ref{fig:STMsummary}(b), which we have assumed to be isotropic. In the other case, $f_{CDW}^3= (\delta_{h} \delta_{k} \delta_{l})$ where $\delta_{h,k,l}$ represent the widths of  the nearest Bragg peak in the three orthogonal reciprocal space directions, shown in Fig.~\ref{fig:-125_1p9K_BraggPeak_fits}(c-e).

Figure \ref{fig:-125_1p9K_BraggPeak_fits} (a-b) shows cuts through the $(-1,2,5)$ Bragg peak. Panels (c-e) show line cuts through the maxima in the $h$, $k$, and $l$, respectively. The peak has a maximum intensity of $1.74 \times 10^{6}$ Hz and widths in the $h$, $k$, and $l$ directions of $0.043 \mathrm{\AA}^{-1}$, $0.032 \mathrm{\AA}^{-1}$, and $0.050 \mathrm{\AA}^{-1}$, respectively. The integrated intensity of this Bragg peak can then be approximated in the same way as the CDW peak shown in Eq.~\ref{eq:integrated_intensity}, $I_{Bragg}^{Integrated} = \frac{1}{2}(I_{Bragg}^{max})(\delta_{h} \delta_{k} \delta_{l}) $, where $\delta_{h,k,l}$ are the measured widths along the crystallographic directions. Using this number, we can place an upper limit on the amplitude charge density of the CDW,
\begin{equation}
    \frac{\rho_{CDW}^2}{\rho_{(-1,2,5)}^2} \lesssim \frac{I_{CDW}^{Integrated}}{I_{Bragg}^{Integrated}},
\end{equation}
where $\rho_{(-1,2,5)}$ is the calculated structure factor of the $(-1,2,5)$ Bragg peak and $\rho_{CDW}$ is the charge amplitude of the CDW.
The structure factor of the $(-1,2,5)$ Bragg peak is calculated to be $\rho_{(-1,2,5)} = 169.38 \: \mathrm{e}^-/v_\mathrm{cell}$, where $v_\mathrm{cell}$ is the volume of the standard orthorhombic unit cell. Comparing this to the amplitude of the charge density of the CDW gives $\rho_{CDW} \lesssim 2.13  \: e^{-}/v_\mathrm{cell}$ in the case that the CDWs have the same width as in STM measurements or $\rho_{CDW} \lesssim 0.70 \: e^{-}/v_\mathrm{cell}$ in the case that the CDWs have the same width as the (-1,2,5) Bragg peak.  

\section{Conclusions}
Our study indicates that UTe$_2$ does not exhibit a bulk CDW of the sort that occurs in many other transition-metal dichalcogenides, such as NbSe$_2$ or TaS$_2$ \cite{Rossnagel2011JPCM}, which exhibit pronounced CDW signatures in both STM and x-ray measurements \cite{Burk1992Science,Chatterjee2015NatComm}. 
Note that, while we have used STM measurements as a guide for where to look in reciprocal space, we performed a 3D momentum sweep covering $>20$ Brillouin zones, and we would have observed the CDW even if its wave vector deviated from what was observed in STM. 

How to interpret the upper bound on the amplitude, $\rho_{CDW}$, depends on whether one assumes the CDW is electronic (the scattering arising purely from a valence modulation) or structural (the scattering arising from a periodic distortion to the crystalline lattice).  

Assuming the CDW is purely structural, the magnitude of the lattice displacement should be of order $u_{max}/a \sim \rho_{CDW}/\rho_{tot}$, where $\rho_{tot}$ represents the total charge density \cite{Rossnagel2011JPCM}. 
Assuming the order is long-ranged (see Section \ref{sec:results}), the amplitude bound $\rho_{CDW} \lesssim 0.70 \: e^{-}/v_\mathrm{cell}$ translates into 
an atomic displacement $u_{max}/a \lesssim 8.93 \times 10^{-4}$, where $a$ is the lattice parameter of UTe$_2$. 
Alternatively, if we now consider the case in which the CDW has a short correlation length, the atomic displacement would be $u_{max}/a \lesssim 2.72 \times 10^{-3}$. 
Both of these numbers are markedly lower than the atomic displacements in other dichalcogenide CDWs, which have distortions $u_{max}/a$ ranging from 1-7\% \cite{Rossnagel2011JPCM}.
We conclude that whatever CDW was observed in STM experiments in UTe$_2$ has a completely different character from CDWs in other dichalcogenides. 

We can also consider the case in which the CDW is assumed to be purely electronic, meaning the atoms do not shift from their equilibrium positions, and the CDW is dominated by a modulation in the valence electron density \cite{Zimmermann1998EPL,Pouget1997SynthMet,Abbamonte2004Nature}.
In this case, our x-ray measurements allow for an amplitude of up to $\sim 2 \: e^{-}$ per unit cell. This would represent a significant fraction of the valence band \cite{2020_Thomas_SciAdv,2022_Aoki_JPSJ,2023_Wilhelm_CommPhys}. Such an amplitude might be comparable to the effect seen in STM, though it would be difficult to reconcile with the absence of a resistive transition in bulk transport measurements in this range of temperatures \cite{Rosa2022CommMatt,2019_Metz_PhysRevB,2022_Sakai_PhysRevMatt}. We cannot, however, completely rule out this possibility. 

A final explanation is that the CDW observed in STM measurements is restricted to the surface of UTe$_2$. This would explain why its presence is so robust in multiple STM measurements \cite{Aishwarya2023Nature,2023_Gu_Nature,Lafleur2024Nature} and absent in transport measurements and the bulk-sensitive x-ray measurements presented here. 
This scenario would be reminiscent of the CDW seen in STM measurements of Na$_{x}$Ca$_{2-x}$CuO$_2$Cl$_2$ \cite{2007_Smadici_PhysRevB}. Such a CDW transition can be explained through the framework of extraordinary phase transitions, in which the critical behavior is different at the surface compared to in the bulk of material \cite{1983_Binder_PhaseTransCritPhen,Brown2005PRB}. As a result the transition temperature of the surface can be higher than that of the bulk. This allows for the possibility that the CDW in UTe$_2$ could be of either electronic or structural origin at the surface, with the absence of a bulk signature in x-ray and transport measurements being a consequence of the small volume fraction of the effect. 
In this scenario, one might expect different surface terminations to lead to different surface ordering effects.

We would like to note that two other reports of experimental investigations of a resonant hard x-ray diffraction experiment \cite{Kengle2024_} and of ultrasound measurements \cite{Theuss2024_} of UTe$_2$ at low temperature, were reported simultaneously with this work. All measurements, performed independently and on samples grown by more than one group, arrive at the same conclusions reported here. 

\section{Acknowledgements}
We thank Priscila F.S. Rosa for helpful discussions. CSK acknowledges useful discussions with Julian May-Mann. This work was primarily
supported by the Center for Quantum Sensing and Quantum Materials, an Energy Frontier Research Center funded by the US Department of Energy (DOE), Office of Science, Basic Energy Sciences (BES), under award no.~DE-SC0021238. 
Growth of UTe$_2$ crystals was supported by National Science Foundation grant NSF-DMR 2105191. STM measurements were supported by DOE BES award no.~DE-SC0022101. Sample characterization was partly supported by DOE BES award no.~DE-SC-0019154 and the Maryland Quantum Materials Center. E.F. was partly supported by NSF grant NSF-DMR 2225920. PA, VM, and JPP gratefully acknowledge additional support from the EPiQS program of the Gordon and Betty Moore Foundation, grant nos. GBMF9452, GBMF9465, and GBMF9071, respectively. 
This research used resources of the Advanced Photon Source, a DOE User Facility operated Argonne National Laboratory under contract no.~DE-AC02-06CH11357.
Identification of commercial equipment does not imply recommendation or endorsement by NIST.

\bibliography{Main-PRB}

\end{document}